\documentclass[aps,superscriptaddress]{revtex4-1}

\usepackage{graphics,graphicx, color}
\usepackage{dcolumn, amsmath,amssymb}

\begin{document}


\title{High-energy neutrinos from multi-body decaying dark matter}


\author{Nagisa Hiroshima}
 \affiliation{Institute for Cosmic Ray Research, The University of Tokyo, 5-1-5 Kashiwanoha, Kashiwa, Chiba 277-8582, Japan}%
 \affiliation{Institute of Particle and Nuclear Studies, Theory Center, KEK,  1-1 Oho, Tsukuba, Ibaraki 305-0801, Japan}
\author{Ryuichiro Kitano}
\affiliation{Institute of Particle and Nuclear Studies, Theory Center, KEK,  1-1 Oho, Tsukuba, Ibaraki 305-0801, Japan}
\affiliation{
The Graduate University for Advanced Studies
(Sokendai), 1-1 Oho, Tsukuba 305-0801, Japan
}%
\author{Kazunori Kohri}
\affiliation{Institute of Particle and Nuclear Studies, Theory Center, KEK,  1-1 Oho, Tsukuba, Ibaraki 305-0801, Japan}
\affiliation{
The Graduate University for Advanced Studies
(Sokendai), 1-1 Oho, Tsukuba 305-0801, Japan
}%
\affiliation{
Rudolf Peierls Centre for Theoretical Physics, The University of Oxford, 1 Keble Road, Oxford OX1 3NP, UK
}
\author{Kohta Murase}
 \affiliation{Department of Physics; Department of Astronomy \& Astrophysics; Center for Particle and Gravitational Astrophysics, The Pennsylvania State University, PA 16802, USA}
 \affiliation{
Yukawa Institute for Theoretical Physics, Kyoto University, Kyoto 606-8502, Japan}


\date{\today}

\begin{abstract}
Since the report of the PeV-TeV neutrinos by the IceCube Collaboration,
various particle physics models have been proposed to explain the
neutrino spectrum by dark matter particles decaying into neutrinos and
other Standard Model particles.
In such scenarios, simultaneous $\gamma$-ray emission is commonly
expected. Therefore, multi-messenger connections are generally important
for the indirect searches of dark matters. The recent development of
$\gamma$-ray astronomy puts stringent constraints on the properties of
dark matter, especially by observations with the {\it Fermi}
$\gamma$-ray satellite in the last several years.
Motivated by the lack of $\gamma$-ray as well as the shape of the
neutrino spectrum observed by IceCube, we discuss a scenario in which
the DM is a PeV scale particle which couples strongly to other invisible
particles and its decay products do not contain a charged particle.
As an example to realize such possibilities, we consider a model of
fermionic dark matter that decays into a neutrino and many invisible
fermions.  The dark matter decay is secluded in the sense that the
emitted products are mostly neutrinos and dark fermions. One remarkable
feature of this model is the resulting broadband neutrino spectra around
the energy scale of the dark matter. We apply this model to multi-PeV
dark matter, and discuss possible observable consequences in light of
the IceCube data. In particular, this model could account for the large
flux at medium energies of $\sim10-100$~TeV, possibly as well as the
second peak at PeV, without violating the stringent $\gamma$-ray
constraints from {\it Fermi} and air-shower experiments such as
CASA-MIA. 
\end{abstract}

\pacs{}

\maketitle

\section{Introduction}
\label{sec:intro}

High-energy neutrinos are powerful probes of the Universe in
astrophysics and cosmology. They enable us to study physical processes
in their sources because they interact with other particles only
through the weak interaction. Searches for astrophysical and
cosmogenic high-energy neutrinos have been drastically advanced in
this decade.  The full IceCube detector was completed in 2010. Since
the first announcement of the detection of high-energy cosmic
neutrinos~\cite{Aartsen:2013bka,Aartsen:2013jdh}, its evidence has
been accumulated~\cite{Aartsen:2014gkd,Aartsen:2014muf,Aartsen:2015ita,Aartsen:2015rwa,Aartsen:2016xlq}.
It is worth noting that some PeV neutrino events have been
detected~\cite{Aartsen:2014gkd,Aartsen:2016xlq}. These neutrinos can
provide us with not only information about the astrophysical
accelerators which have abilities to generate cosmic rays with
${\cal O}(10-100)$~PeV, but also some hints about physics beyond the
Standard Model.

There are numbers of interpretations on the origin of the TeV-PeV
neutrinos (see a review~\cite{Halzen:2016gng}).  So far no clustering in time and/or space has been
seen. The observed diffuse neutrino intensity above $\sim0.1$~PeV is
$E_\nu^2\Phi_\nu \sim3\times{10}^{-8}~{\rm GeV}~{\rm cm}^{-2}~{\rm s}^{-1}~{\rm sr}^{-1}$
for the sum of three flavors, and the spectrum can be approximately
described by a single power law with its spectral index, $s_{\nu}\simeq2.0-2.7$ up to ${\cal O}(1)$
PeV~\cite{Aartsen:2014muf,Aartsen:2015rwa,Aartsen:2015zva,Aartsen:2016xlq}.
Remarkably, the latest analysis focusing on medium-energy starting
events below 0.1~PeV~\cite{Aartsen:2014muf,Chen:2014gxa,Aartsen:2015ita,Vincent:2016nut,Palladino:2016xsy}
as well as the conventional shower analysis~\cite{Aartsen:2015ita,Aartsen:2015zva} have indicated that
the neutrino flux in the 30~TeV range is as high as $E_\nu^2\Phi_\nu \sim{10}^{-7}~{\rm GeV}~{\rm cm}^{-2}~{\rm s}^{-1}~{\rm sr}^{-1}$.
Consequently, for a power-law fitting, the medium-energy spectral
index below $\sim0.1$~PeV is $s_{\rm ME} \sim2.5-2.7$, which suffers
from a $\sim3\sigma$ tension with the high-energy spectral index,
$s_{\rm HE}\sim2.0-2.2$~\cite{Aartsen:2016xlq}.  The flavor ratio is consistent with
$\nu_e:\nu_\mu:\nu_\tau\approx1:1:1$~\cite{Aartsen:2015ivb,Aartsen:2015ita},
as expected in the long baseline limit of neutrino oscillation.
  
In astrophysical models, those high-energy neutrinos can be produced by proton-proton ($pp$) and/or proton-gamma ($p\gamma$) interactions~\cite{Murase:2013rfa,Winter:2013cla,Murase:2015xka} induced by accelerated primary cosmic rays.
In this kind of scenarios, extragalactic sources may be favored because the observed neutrinos are isotropically distributed. In particular, cosmic-ray reservoirs such as starburst galaxies and galaxy clusters give a grand-unified picture of the observed neutrinos, $\gamma$ rays, and ultra-high energy cosmic rays~\cite{Murase:2013rfa,Murase:2016gly,Fang:2017zjf}.
On the other hand, important constraints on astrophysical models for IceCube's neutrinos have been placed by the diffuse isotropic $\gamma$-ray background (IGRB), which is measured by the {\it Fermi} $\gamma$-ray satellite~\cite{Ackermann:2014usa}. In either $pp$ or $p\gamma$ process, $\gamma$ rays should be commonly generated and their fluxes are known to trace the neutrino flux approximately in a model-independent way. This fact inevitably induces a severe problem that those $\gamma$-rays may exceed the measured IGRB flux at around $10-100$~GeV~\cite{Murase:2013rfa}. 
This is especially the case if the neutrino flux excess at medium energies is real, and hidden cosmic-ray accelerators, which are opaque to $\gamma$ rays, are needed to reconcile the neutrino and $\gamma$-ray data~\cite{Murase:2015xka}. 

Not only astrophysical interpretations but also the new physics models are feasible at present. In particular, high-energy neutrinos can be emitted through decays of a long-lived heavy particle with a mass of ${\cal O}(1)$--${\cal O}(10)$
PeV~\cite{Feldstein:2013kka,Esmaili:2013gha,Bai:2013nga,Bhattacharya:2014vwa,Higaki:2014dwa,Rott:2014kfa,Fong:2014bsa,Esmaili:2014rma,Dudas:2014bca,Murase:2015gea,Fiorentin:2016avj,Dev:2016qbd,DiBari:2016guw,Cohen:2016uyg,Borah:2017xgm,Chianese:2016opp,Chianese:2016kpu,Chianese:2016smc,Bhattacharya:2016tma,Ko:2015nma}, of which the dark matter (DM) is expected to consist. 
Using the IceCube data, lower bounds on the lifetimes of PeV or heavier DMs have been obtained, $\tau\gtrsim10^{27}-10^{28}$~sec~\cite{Esmaili:2012us,Murase:2012xs,Rott:2014kfa}, which should be obviously longer than the cosmic age $\sim 10^{18}$~sec~\cite{PalomaresRuiz:2007ry}.  
Among the theoretical particle physics models of the DM scenarios, some groups have considered a DM particle decaying into two-body and/or multi-body daughter particles which produces line and broad spectra of the high-energy neutrinos in order to fit the data~\cite{Feldstein:2013kka,Higaki:2014dwa,Fong:2014bsa,Dudas:2014bca,Boucenna:2015tra} (see also Refs.~\cite{Ema:2013nda,Ema:2014ufa,Ema:2016zzu} for alternative models).

However, $\gamma$-ray constraints are crucial for many DM models. There are numbers of DM models  that include $\gamma$ rays, charged leptons, quarks and so on as the final states of the decaying DM.  
Whatever the decay products are, the resulting $\gamma$-ray emission should not violate the observed diffuse $\gamma$-ray flux. For example, charged leptons can easily be converted into secondary $\gamma$-rays at lower energies, e.g., through the inverse Compton scattering off the background photons, as is the case with astrophysical situations. Quarks also emit $\gamma$ rays because they hadronize during their propagation, then, fragment into a lot of pions that immediately decay into a broad spectrum of $\gamma$-rays, charged leptons, and neutrinos.
As a result, stringent constraints are placed by the {\it Fermi}
$\gamma$-ray data~\cite{Murase:2015gea,Cohen:2016uyg} as well as
non-observations of diffuse sub-PeV $\gamma$-rays in air-shower
experiments such as KASCADE and 
CASA-MIA~\cite{Ahlers:2013xia,Kalashev:2014vra,Murase:2015gea,Esmaili:2015xpa,Kalashev:2016cre}. In
particular, Ref.~\cite{Cohen:2016uyg} performed thorough analyses on the
heavy DM that decays into all final states of Standard Model
particle-antiparticle pairs via flavor-conserving two-body decay. It is
found that neutrinophillic DM needs to avoid the $\gamma$-ray
constraints, which is especially the case if the medium-energy excess in
the IceCube neutrino flux~\cite{Aartsen:2014muf,Chen:2014gxa,Aartsen:2015ita,Vincent:2016nut,Aartsen:2016xlq}
is attributed to decaying DM that gives comparable contributions to
Galactic and extragalactic components~\cite{Murase:2015gea,Chianese:2016opp,Chianese:2016kpu,Chianese:2016smc,Cohen:2016uyg}.

The required mass of the DM, a few PeV, also has some tension with the
standard thermal relic abundance when we assume that the DM to be an
elementary particle. The unitarity of the annihilation cross section put
an upper limit on the mass of the DM to be about a hundred
TeV~\cite{Griest:1989wd}, which is an order of magnitude smaller. In
addition, the energy spectrum measured at IceCube has a board component
around ${\cal O}(10-100)$~TeV as well as a peak at PeV. This structure
does not seem to be easily explained solely by the decay of the DM.

Under these circumstances, we propose a new decaying DM scenario where
the DM is very weakly coupled to the Standard Model particles so that it
is almost stable, but is very strongly coupled to other
non-Standard-Model particles so that it has a finite size (form factor)
to make the annihilation cross section larger than the naive unitarity
limit and also to make the broad spectrum of neutrinos possible in
addition to the line at PeV.
In order to make the discussion concrete, we set up an effective
particle physics model where the DM particle interact perturbatively to
other particles and thus one can evaluate the shape of the neutrino
spectrum.
In this scenario, DM decays to neutrinos with few $\gamma$-rays in two
branches; in one branch, a fermionic DM particle decays into ${\cal
O}$(10) particles including a neutrino. A broad spectrum of the produced
neutrino is predicted in this branch. In the other branch, the DM
undergoes two-body decay into a neutrino and another invisible particle
which leads to a line-like feature of the neutrino spectrum. Negligible
amount of photons and charged leptons are produced in this model. This
property of the DM leads to line and broad spectra of high-energy
neutrinos.
%
If the branching fractions into multi-body and two-body are comparable,
one could simultaneously explain both of the neutrino flux around the
cutoff of a few PeV and a peak structure below ${\cal O}(1)$ PeV with
satisfying the constraint from the $\gamma$-ray observations.
For other mechanisms to fit the complex features of the neutrino spectrum, see ~\cite{Ioka:2014kca,Ng:2014pca,Blum:2014ewa,Ibe:2014pja,Araki:2014ona,Shoemaker:2015qul} by neutrino absorptions, or ~\cite{Bustamante:2016ciw,Pagliaroli:2015rca,Shoemaker:2015qul} by neutrino decays. 

The outline of this paper is as follows. 
The particle physics model of DM is introduced in Sec.~\ref{sec:model}. 
In Sec.~\ref{sec:astrophysics}, we explain both astrophysical and cosmological effects on high-energy neutrinos. 
In Sec.~\ref{sec:results}, we show our results. 
Finally, Sec.~\ref{sec:conclusion} is devoted to summarize and conclude our findings in this paper.

\section{The Model}
\subsection{Particle physics models}
\label{sec:model}

The data from the IceCube experiment tells us that there is certainly an unexplored source of high-energy neutrinos in the sky. 
One important fact in this discussion is that the source of the high energy neutrinos should not emit too many $\gamma$-rays in order to be consistent with the observation of the IGRB. This is a non-trivial constraint for both the astrophysical and DM models.

In this section, we consider a model of decaying DM in which the DM
mainly decays into neutrinos but not to charged leptons so that the
emission of $\gamma$ rays is highly suppressed.
Since the neutrino and the charged lepton form a doublet of the $SU(2)$
weak interactions, $\ell$, in the Standard Model, the DM cannot be a
gauge singlet that would not distinguish the neutrinos and charged
leptons unless we include the right-handed neutrinos as the final
state~\cite{Fong:2014bsa,Dudas:2014bca}.
%
%
The simplest possibility to achieve the absence of the charged lepton
mode is that the DM is a neutral component $N^0$ of an $SU(2)$ doublet,
$L$,
\begin{eqnarray}
  \label{eq:doublet}
  L = \left(
  \begin{array}{ll}
    N^0 \\
    E^-
  \end{array}
   \right),
\end{eqnarray}
and couples to the lepton doublet,
$l$, as
\begin{eqnarray}
  \label{eq:lagrange1}
  {\cal L}_{X} =
- m_{\rm DM} \bar L L + 
\left(
   \epsilon  \overline{L} \ell X + {\rm h.c.}
\right),
\end{eqnarray}
in the Lagrangian. Here we introduced the DM sector particles, $L$ and $X$, respectively as
a Dirac fermion with hypercharge $-1/2$ and a gauge-singlet real scalar
field.
In the absence of the second term, the lighter of $N^0$ and $E^-$ is stable since there is no decay channel. A dimensionless parameter
$\epsilon$ characterizes the coupling between the DM sector particles and the Standard Model ones, and we assume that is somehow very much suppressed so that $N^0$ has survived today as DM.
The mass of the DM is assumed to be $m_{\rm DM} = {\cal O}({\rm PeV})$
to explain the PeV neutrinos detected at the IceCube as the primary
decay products. The particle $X$ is in general massive, but we ignore
its mass compared to the PeV energy scale in the following discussion
for simplicity. 
%
Since the interactions between $L$ and the Higgs field in the Standard
Model is suppressed by assumption, the mass splitting between $N^0$ and
$E^-$ is mainly through the radiative correction, which makes $E^-$ a
little bit heavier than $N^0$ by ${\cal O}(\alpha m_W/(4\pi))\sim 300$~MeV~\cite{Nagata:2014wma}.
We suppressed the flavor indices of $\ell$ for simplicity. The flavor
content of the high energy neutrinos can be arranged arbitrarily by the
choice of the flavor dependence of the $\epsilon$ parameter.

We further assume that $X$ promptly decays into some other particle $S$'s
in the DM sector through the following interaction terms:
\begin{eqnarray}
  \label{eq:lagrange1-2}
  {\cal L}_{\rm int} = {\cal L}_X
+  \frac{1}{M^{3n-3}} X S^{2 n} 
+  \frac{1}{M_*^{3n-1}} \overline{L} \ell S^{2 n}
+ {\rm h.c.}
\end{eqnarray}
Here $S$ denotes a massless Weyl spinor field. (The contraction of the spinor and $SU(2)$ indices are implicit.) By assuming a discrete
$Z_{2n}$ symmetry, one can arrange the $S$ particle to be massless and the above operators to be the lowest dimensional ones.
The third term is allowed (or induced) in general in the presence of the second term in eq.~\eqref{eq:lagrange1} and the second term in
eq.~\eqref{eq:lagrange1-2}.
We discuss a possible theoretical background behind this model in Appendix~\ref{sec:setup}.

Through those interaction terms, the decaying $N^0$ simultaneously produces the line ($N^0 \to \nu + X$) and the broad spectra ($N^0 \to \nu + 2 n S$) of the active neutrinos $\nu$ if $n$ is large enough.
The branching fractions into $X$ and $2 n S$ can be comparable in a model
with a strongly coupled DM sector. (See Appendix~\ref{sec:setup}.)
Although, in a strongly coupled theory, it is possible to add any higher
dimensional operators (or equivalently a form factor) to make the shape
of the spectrum arbitrary, we will use a standard description of the
decay of weakly coupled particles below in order to make discussions
concrete and to get a sense of feasibility to fit the broad spectrum of
$0.01-1$~PeV neutrinos by the decay of DM.

The lifetime of the DM particle is estimated to be 
\begin{align}
 \Gamma_{\rm line} \sim {\epsilon^2 \over 16 \pi}  m_{\rm DM}.
\end{align}
In order to explain the flux of the PeV-energy line, one needs $\epsilon
\sim 10^{-29}$. 

If one also tries to fit the broad spectrum at $0.01-1$~PeV, it is required that $2nS$ is one of the main decay mode. The size of $M_*$, in this case, depends on $n$ as follows.
%
By using the third term of (\ref{eq:lagrange1-2}), the decay rate (or partial decay width) into the broad $\nu$ mode is approximately
estimated to be
\begin{eqnarray}
  \label{eq:decayRate}
  \Gamma_{\rm broad} &\sim& \frac{1}{16\pi} 
  \left[ \frac1{(4\pi)^2}  \right]^{2n-1}
  \left( \frac{m_{\rm DM}}{M_*}  \right)^{2(3n-1)} m_{\rm DM}, 
\end{eqnarray}
The partial widths $\Gamma_{\rm line}$ and $\Gamma_{\rm broad}$ become comparable when
\begin{align}
 \epsilon \sim 
\left(
1 \over 4 \pi
\right)^{2n-1}
\left(
{m_{\rm DM} \over M_*}
\right)^{3n-1},
\end{align}
which is the case when $L$, $X$ and $S$ are strongly coupled as we can see in Appendix~\ref{sec:setup}.

Here, we briefly discuss how the correct amount of $N^0$ is produced in
the early Universe.
Since $N^0$ interacts with the Standard Model particles through the
electroweak interactions, the DM sector particles are thermally produced
as long as the reheating temperature after inflation exceeds ${\cal
O}({\rm PeV})$.
In the standard freeze-out scenario, when the temperature drops below
PeV, the pair annihilation process of $N^0$ into $X$ or $S$ reduces the
number density of $N^0$. The relic abundance of the PeV mass particle
$N^0$, however, is larger than the observed one, provided the
annihilation cross section is within the unitarity
limit~\cite{Griest:1989wd} that puts a upper bound on the DM mass to be
approximately a hundred TeV.
One simple possibility to reconcile the DM abundance is to assume that
the annihilation cross section goes beyond the unitarity limit, which
means $N^0$ is a composite particle with a finite size rather than an elementary particle.
The required size is $r \sim  6\times10^{-19}$cm for $\sigma v\sim \pi r^2 \sim3\times10^{-26}$cm$^3$/s. This is interestingly
the size expected from the naive dimensional analysis, $r \sim 4 \pi /
m_{\rm DM}$ for $m_{\rm DM}\sim$0.4PeV. It is somewhat interesting to note that the mass and the spectrum both
point to the strongly coupled nature of the DM.

The violation of the unitarity limit does not means that the naive dimensional analysis overestimates the cross sections. The unitarity is maintained for each partial waves and adding up them provides the consistent estimates~\cite{Griest:1989wd}. The estimate of the scattering amplitude in the naive dimensional analysis is based on the assumption that the perturbative expansion breaks down; all levels in the perturbative expansion give contributions of the same order of magnitude. This is known to give good estimates in the low-energy hadron physics.
The scattering amplitudes $|{\cal M}|$ are estimated to be of order $|{\cal M}| \sim(4\pi)^2$ in the naive dimensional analysis while the unitarity limit for each partial wave is $|{\cal M}|\lesssim{\cal O}(4\pi)$. Therefore, the annihilation cross section of our DM becomes two orders of magnitude higher than that of the naive unitarity limit.
For an example of models to go further beyond the unitarity limit, see
Ref.~\cite{Harigaya:2016nlg}.
Another possibility is to assume a dilution of DM by a late-time entropy 
production due to, for example, a decay of a scalar condensation~\cite{Scherrer:1984fd}.
Yet another possibility would be the scenario with a low reheating
temperature. The production of such heavy DM
has been shown to be possible in the previous literature~(e.g.,~\cite{Chung:1998ua,Chung:1998zb}), and 
the detailed cosmological scenarios will be
discussed in a separate paper.

We also mention the cosmological history of the heavy charged lepton
$E^-$. It is natural that $E^-$ has the same abundance as $N^0$ in the
early Universe since they are the same particle before the electroweak
phase transition. The mass difference between $N^0$ and $E^-$ would be
expected to be the order of $\Delta m \sim \alpha m_W/(4\pi)$ with the
weak boson mass $m_W$, which gives ${\cal O}(\Delta m)\sim 300$~MeV.  Then the
decay width of $E^-$ is estimated to be $\Gamma_{E^-} \sim G_{\rm F}^2
\Delta m^5 \sim (10^{-7}~{\rm sec})^{-1}$.  This means that $E^-$ had
disappeared before the beginning of Big-bang nucleosynthesis. 

About cosmological histories of $X$ and $S$, it is expected that $X$ had decayed completely into $S$'s in a short time. 
The thermalized $S$ around the energy scale of PeV are diluted by a late-time entropy production including the one after the QCD phase transition. In this case, we predict a dark radiation component by the relic abundance of $S$ as an effective number of neutrino species $N_{\rm eff}$ to be of the order of 0.1, which will be tested by future observations, e.g., through precise CMB and 21cm line observations~\cite{Kohri:2013mxa}.

In a more general setup, only a fraction of DM may consist of $N^0$ by
the ratio of $N^0$ to the total DM density,
$f_{N^0} = \Omega_{N^0}/ \Omega_{\rm DM}$ which ranges $f_{N^0}$= 0 --
1 with $\Omega_i$ the cosmological $\Omega$ parameter of the
$i$-particle. Then, a flux of daughter particles produced by the
decaying $N^0$ is scaled by a factor of $f_{N^0}$. In this situation,
hereafter we take this notation as read even if it is not stated
explicitly.

\subsection{Neutrino Spectra}
\label{sec:astrophysics}

As mentioned in Sec.~\ref{sec:model}, we assume the DM particle $N^0$ mainly decays into a neutrino $\nu$ and $2n$ fermion particles $S$:
\begin{equation}
N^0\to\nu+2nS
\end{equation}
Hereafter, we use a positive integer $N \equiv{3n} +1$ instead of $n$.
In the massless limit of $\nu$ and $S$, the distribution function of the neutrino can be written as
\begin{equation}
f(x)=\frac{1}{\Gamma}\frac{d\Gamma}{dx}=4N(N-1)(N-2)\cdot{x^2}(1-2x)^{N-3}
\end{equation}
with
\begin{equation}
x=E/m_\mathrm{DM} \ \ \ \ \ \ \ \ (0\leq{x}\leq\frac{1}{2}),
\end{equation}
where $\Gamma=\tau^{-1}$ is the total decay width of the DM with its
lifetime $\tau$. The distribution function $f(x)$ is normalized so
that $\int{f(x)dx}=1$.

We also consider a mode in which the DM particle decays into two
particles including a neutrino. In this case, each particle
approximately has the energy equal to a half of the DM mass. While the
multi-body decay of the DM produces the broad spectrum of the neutrino,
this two-body decay leads to a line spectrum. The branching ratio of
each mode is,
\begin{equation}
\mathrm{BR}_i=\Gamma_i/\Gamma=\frac{\Gamma_i}{\Gamma_\mathrm{line}+\Gamma_\mathrm{broad}} \ \ \ \  \ \ \ i=\mathrm{line \ \ or \ \  broad}.
\end{equation}
The indices $i$ = ``line'' and ``broad'' mean the two-body and multi-body decay channels of the DM, respectively.

For a given particle physics model, one can calculate neutrino spectra as follows. We consider the late-time decay of the heavy DM, where both extragalactic and Galactic contributions are relevant.  
The differential flux per energy, area, time, and solid angle, of the extragalactic component is given by~(e.g.,~\cite{Murase:2012xs} and references therein) 
\begin{equation}
\label{eq:DefnuFnu}
\Phi^{\rm EG}(E)= \frac{1}{4 \pi H_0} \int dz \,\,\, \frac{1}{\sqrt{\Omega_\Lambda+{(1+z)}^3 \Omega_m}} \,\,\, 
\frac{\bar{\rho}_{\rm DM}}{m_{\rm DM} \tau} \frac{d S}{d E'},
\end{equation}
where $E'=(1+z)E$, $dS/dE'=f(x')/m_{\rm DM}$ is the primary spectrum, and $\bar{\rho}_{\rm DM}$ is the DM energy density in the Universe. We adopt $H_0 \equiv 100h =70.2~{\rm km}~{\rm s}^{-1}~{\rm Mpc}^{-1}$, $\Omega_{\rm dm}=0.229$, $\Omega_m=0.275$ and $\Omega_{\Lambda}=0.725$~\cite{Komatsu:2010fb} but the results are insensitive to small changes in the cosmological paramaters.  

The Galactic component is given by
\begin{eqnarray}
\Phi^{\rm G} (E;\psi) = \frac{R_{\rm sc} \rho_{\rm sc}}{4 \pi m_{\rm DM} \tau} \frac{d S}{dE}  {\mathcal J} (\psi),
\end{eqnarray}
where the $\mathcal J$ factor is
\begin{equation}
{\mathcal J}(\psi)= \frac{1}{R_{\rm sc} \rho_{\rm sc}} \int_0^{l_{\rm max}} dl \,\,\, \rho_{\rm DM} (r), 
\end{equation}
where $\rho_{\rm DM}(r)$ is the DM density profile in the Milky Way, and we use the Navarro-Frenk-White (NFW) profile for the calculations~\cite{Navarro:1996gj}. In the case of decaying DM, contrary to DM annihilation, adopting other profiles do not affect our results significantly. The spatial distribution is nearly isotropic but there is a large-scale anisotropy~\cite{Ahlers:2013xia,Bai:2014kba,Murase:2015gea,Denton:2017csz}. For the purpose of this work, it is enough to use the average ${\mathcal J}$ factor in a cone with half-angle $\psi$ around the Galactic center:
\begin{eqnarray}
{\mathcal J}_{\Omega} = \frac{2 \pi}{\Omega} \int_{\cos \psi}^1 d (\cos \psi') \,\,\, {\mathcal J} (\psi')
\end{eqnarray}
where $\Omega=2 \pi (1-\cos \psi)$ is the solid angle of a field of view. We use $\rho_{\rm sc}=0.3~{\rm GeV}~{\rm cm}^{-3}$, $R_{\rm sc}=8.5$~kpc, and $\psi=\pi$ throughout this work.

We include those cosmological modification in the current analyses.  However, it is remarkable that approximately it has a peak at $x_{\rm peak} = E_{\rm peak}/m_{\rm DM}= 2/(N+1)$ to be
\begin{equation}
\Phi(E) \left|_\mathrm{peak}\right. \approx
5.7\times10^{-8}
\left(\frac{A(N)}{0.04}\right)
\left(\frac{\tau}{10^{27}~\mathrm{s}}\right)^{-1}
\left(\frac{\mathrm{BR}_i}{1}\right)
\left(\frac{f_{N^0}}{1}\right)
\left[\frac{1+1.6({\mathcal J}_{\Omega}/2)}{2.6}\right]
~\mathrm{GeV}{\rm cm}^{-2}\mathrm{s}^{-1}{\rm sr}^{-1},
\label{analytic}
\end{equation}
with 
\begin{equation}
A(N)=\frac{2^6 N(N-1)(N-2)}
     {\left({N+1}\right)^4}
     \left(\frac{N-3}{N+1}\right)^{N-3}.
\end{equation}

 
\section{Results}
\label{sec:results}
In this paper, by adopting the above setups for the emission mechanisms of high-energy neutrinos, we study the following two scenarios,
\begin{enumerate}
\item Pure DM scenario
\item Hybrid scenario with DM and astrophysical contributions
\end{enumerate}
in the next subsections.

We compare our results with the IceCube data in two ways: the
deposit energy distribution of high-energy starting event (HESE) neutrinos~\cite{Aartsen:2014gkd} 
and the all-flavor neutrino spectrum obtained by the combined likelihood analysis~\cite{Aartsen:2015ita}. 
The HESE data are composed of high-energy showers and some track samples, which are less
contaminated by the atmospheric backgrounds and describe neutrinos
above $\sim60$~TeV. We first use the four-year HESE data in this work,
and calculate the deposit-energy distribution following
Ref.~\cite{Blum:2014ewa}. This is more appropriate to see whether a
line spectrum is compatible with the data or not.  We also use the
latter for the specific motivation to discuss the origin of the
medium-energy flux excess at around 30~TeV. Atmospheric neutrinos are
more important below $\sim100$~TeV, and more sophisticated analyses
with a veto have been performed as well as conventional shower
analyses~\cite{Aartsen:2014muf,Aartsen:2015zva}. In this case, we need
more dedicated analyses taking into account the veto efficiency, which
is beyond the scope of work. For simplicity, we show the neutrino spectrum data
obtained by the combined likelihood analysis~\cite{Aartsen:2015ita}. This is
justified since in our model, only the broad component from multi-body
decay could account for the 10-100 TeV data.


\subsection{Pure DM contributions}
\label{sec:pureDMfit}
First, we consider the case that there is no astrophysical contribution
to the observed spectrum of the high-energy neutrinos. 
We try to fit whole part of the spectrum observed by the IceCube by the
line and broad components originated from the decaying DM.

\subsubsection{Model 1:}
\begin{figure*}[tbhp]
\centering 
\includegraphics[width=0.65\textwidth,trim=20 200 20 150, clip]{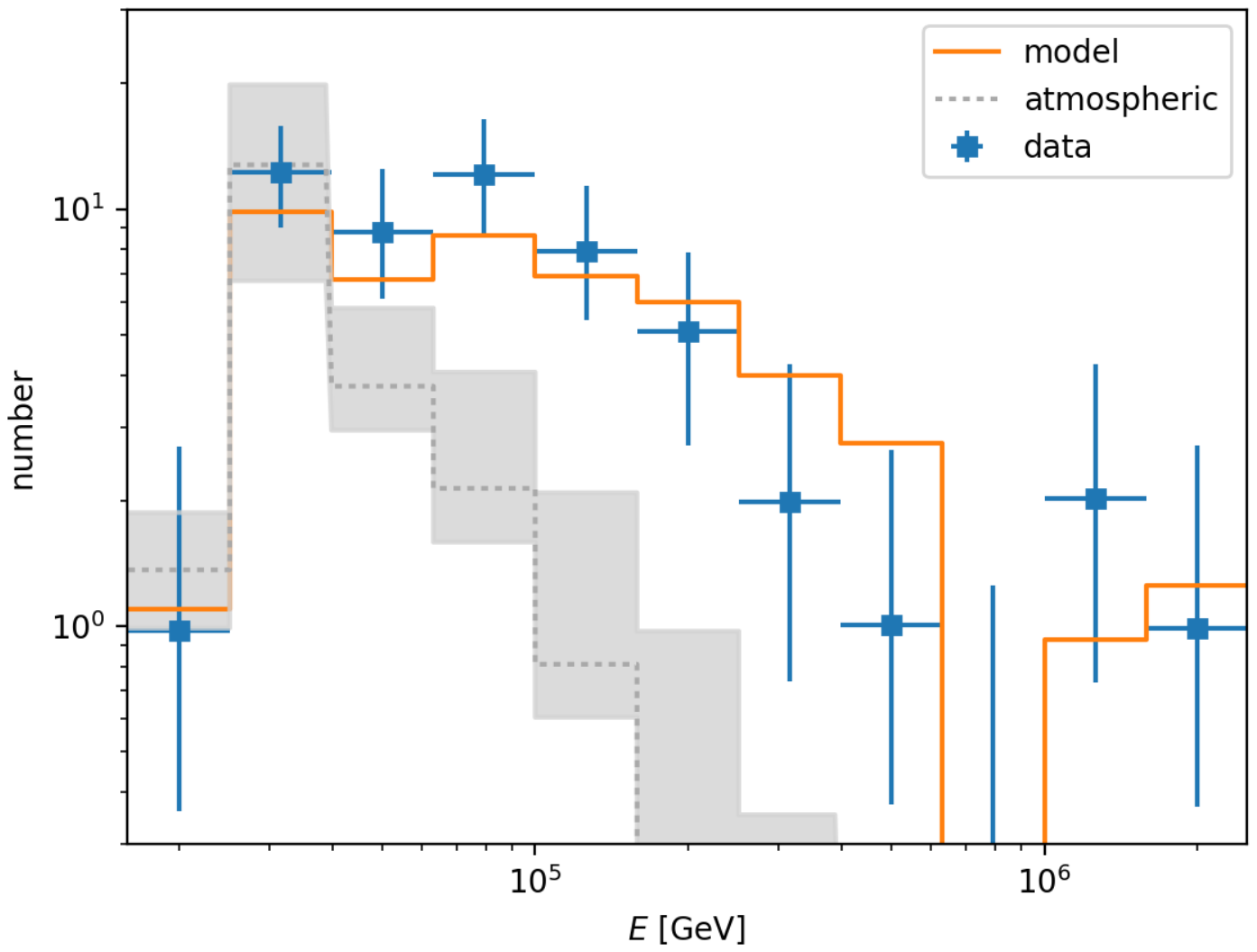}
\caption{\label{pureDMdepo} 
  Neutrino deposited energy histogram predicted by a decaying DM without
  assuming astrophysical contributions (Model 1) . The dots with error bars represent
  the observed data points. The solid line is the theoretical
  prediction in the model. We assume that the DM with its mass
  $m_\mathrm{DM}=4$ PeV decays into $N\sim30$ particles. The branching
  ratio into the two-body decay is taken to be
  BR$_\mathrm{line}$=0.034. In this model, the lifetime is fitted to
  be $1.46\times10^{27}$sec. We also show the atmospheric contribution with short-dashed lines with its uncertainty  (gray shaded regions).
  }
\centering 
\includegraphics[width=0.65\textwidth,trim=20 200 20 150, clip]{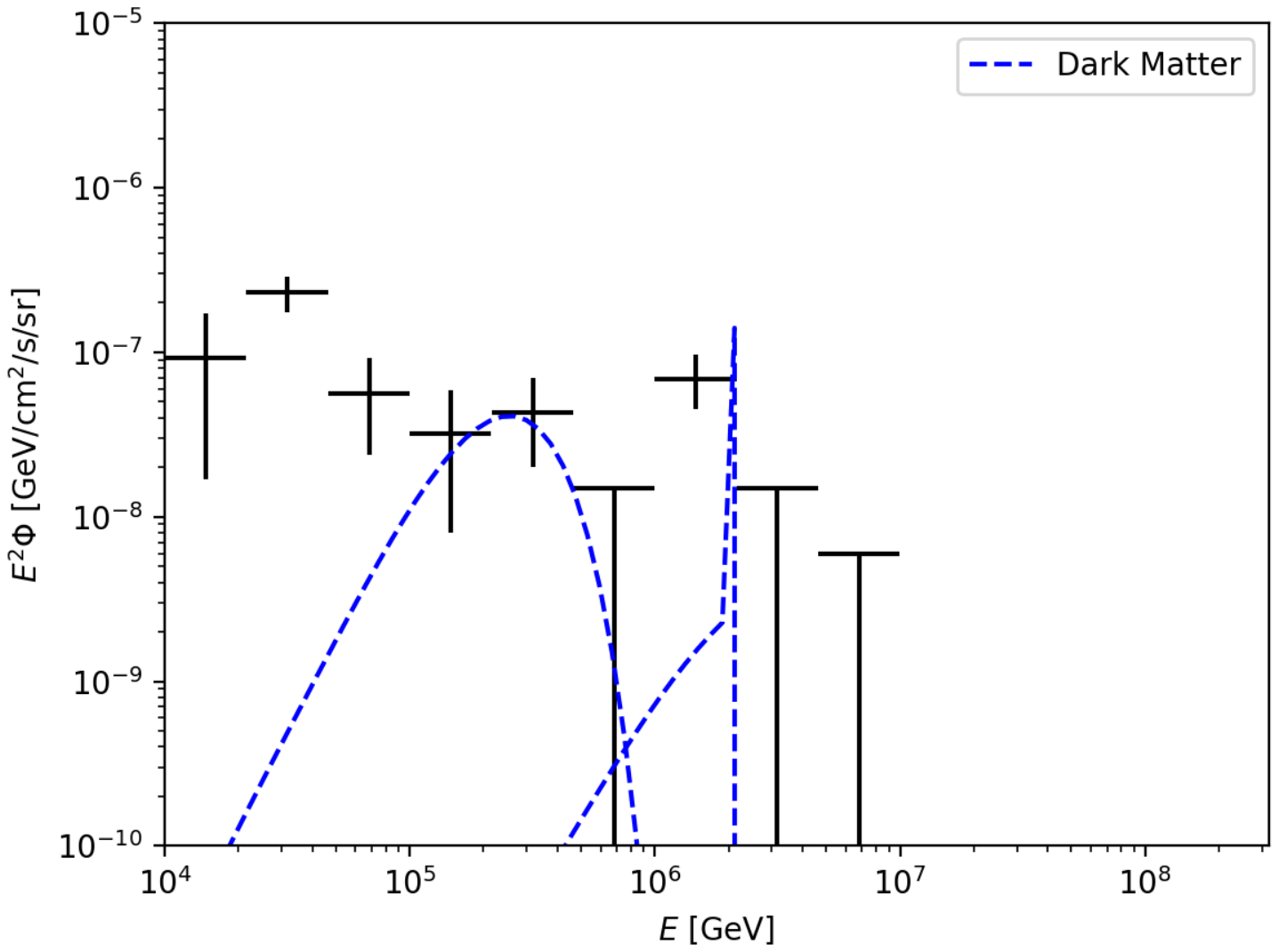}
\caption{\label{pureDMsou} Neutrino source spectrum in Model 1 . Black points with error bars are data. Dashed line represents the model prediction. Model parameters are same in those used in Fig.~\ref{pureDMdepo}. 
The IceCube data for the sum of all flavors come from the results of the combined likelihood analysis.
}
\end{figure*}

We show in Fig.~\ref{pureDMdepo} and Fig.~\ref{pureDMsou} the neutrino
deposited energy and the source spectrum, respectively, with the
parameters, the DM mass, $m_\mathrm{DM}=4$~PeV, the number of daughter
particles, $N\sim30$ ($n=10$), the lifetime $\tau=1.46\times10^{27}$~sec, and
the branching fraction into the two-body decay, BR$_\mathrm{line}=0.034$.
Although we have not performed quantitative analyses that are beyond the scope of this work, 
we see a reasonable agreement with the observed spectrum in Fig.~\ref{pureDMdepo} for 
a wide range of deposited energies, 10~TeV$-2$~PeV, that includes both the line and broad spectrum.

On the other hand, there is some amount of deficit in the lower energy bins around 10-100TeV in Fig.~\ref{pureDMsou} even for a very large number of $N$.
In order to explain the neutrino source spectrum including the medium energies, we may need to assume either additional component such as astrophysical
sources or some level of contaminations by the atmospheric muons.


\subsection{DM + astrophysical contributions}


Motivated by the results in Model 1, we consider cases where an additional astrophysical component contributes to the spectrum.
In such a two-component scenario~\cite{Chen:2014gxa}, we have two possibilities: the astrophysical component for 10-100~TeV neutrinos and the DM component for $\gtrsim100$~TeV neutrinos, or vice versa.
We first discuss the possibility of filling the deficit in the lower-energy bins by astrophysical sources (Model 2a). 
Another possibility of having astrophysical component all the way up to PeV is also examined as
Model 2b. 
%
Model 2b is motivated by a non-trivial success of an
astrophysical model. The observed high-energy neutrino flux is
remarkably consistent with the Waxman-Bahcall bound with
$E_\nu^2\Phi_\nu\sim3\times{10}^{-8}~{\rm GeV}~{\rm cm}^{-2}~{\rm
  s}^{-1}~{\rm sr}^{-1}$~\cite{Waxman:1998yy}.
Astrophysical components are expected to have a power-law spectrum of
high-energy neutrinos, and a flat energy spectrum of $E_\nu^2\Phi_\nu$
can explain not only the IceCube data above 0.1~PeV but also the
$\gamma$-ray and ultra-high energy cosmic-ray data~\cite{Murase:2016gly}.


For simplicity, we here adopt a power-law spectrum with an exponential
cutoff, $\Phi_{\nu}^{\rm astro}\propto E^{-s_{\rm astro}} \exp(-E/E_{\rm
cut})$, for an astrophysical component. The spectral index of the
astrophysical neutrinos is fixed to $s_{\rm astro} = 2.0$.
The cutoff scale is set to ${\cal O}$(10) TeV (model 2a) and
${\cal O}$(1) PeV (model 2b), respectively.  In the former model, the
DM is used for an explanation of the high-energy data, while the
medium-energy data are explained by astrophysical sources (e.g.,
hidden cosmic-ray accelerators). In the latter model, the
astrophysical component explains the data above 0.1~PeV, while the
medium-energy data are attributed to the DM.

\subsubsection{Model 2a: Cutoff of the astrophysical components at ${\cal O}$(10) TeV}

\begin{figure*}[htbp]
\centering 
\includegraphics[width=0.65\textwidth,trim=20 200 20 150, clip]{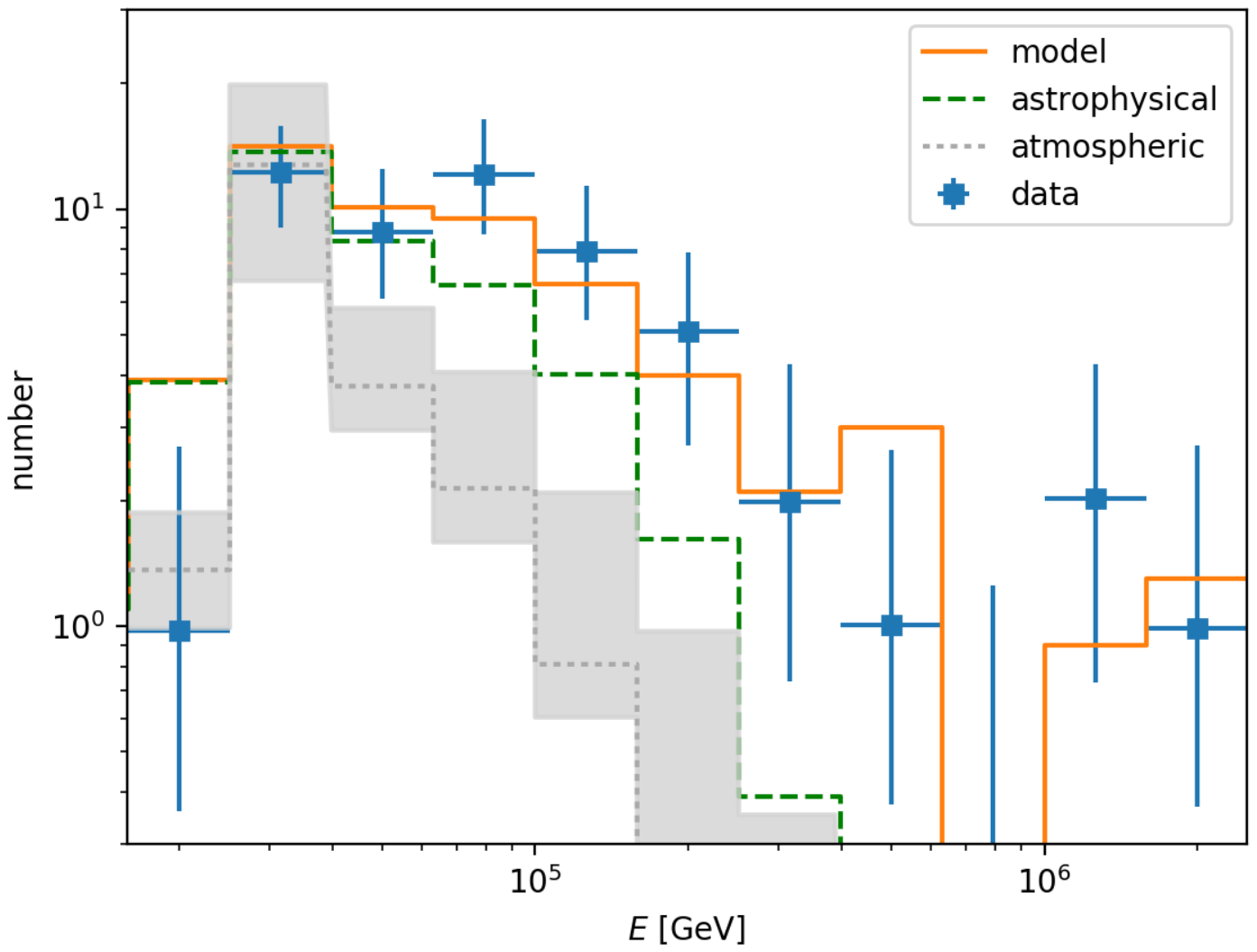}
\caption{\label{astroDM2depo} Deposited energy histogram of the neutrino spectrum combined the astrophysical component with those of the two- and multi-body decaying DM contributions (Model 2a).
The total (astrophysical)
  contribution is represented in the solid (long-dashed) line. The short-dashed and shaded region corresponds to the atmospheric contributions and its uncertainty, which is same as those in Fig.~\ref{pureDMdepo}. In
  this case, the DM with its mass $m_\mathrm{DM}=4$ PeV also decays into
  $N\sim30$ particles. The branching ratio into the line spectrum and the
  lifetime is assumed to be BR$_\mathrm{line}=0.080$, and
  $\tau=3.41\times10^{27}$~sec, respectively.}
\includegraphics[width=0.65\textwidth,trim=20 200 20 150, clip]{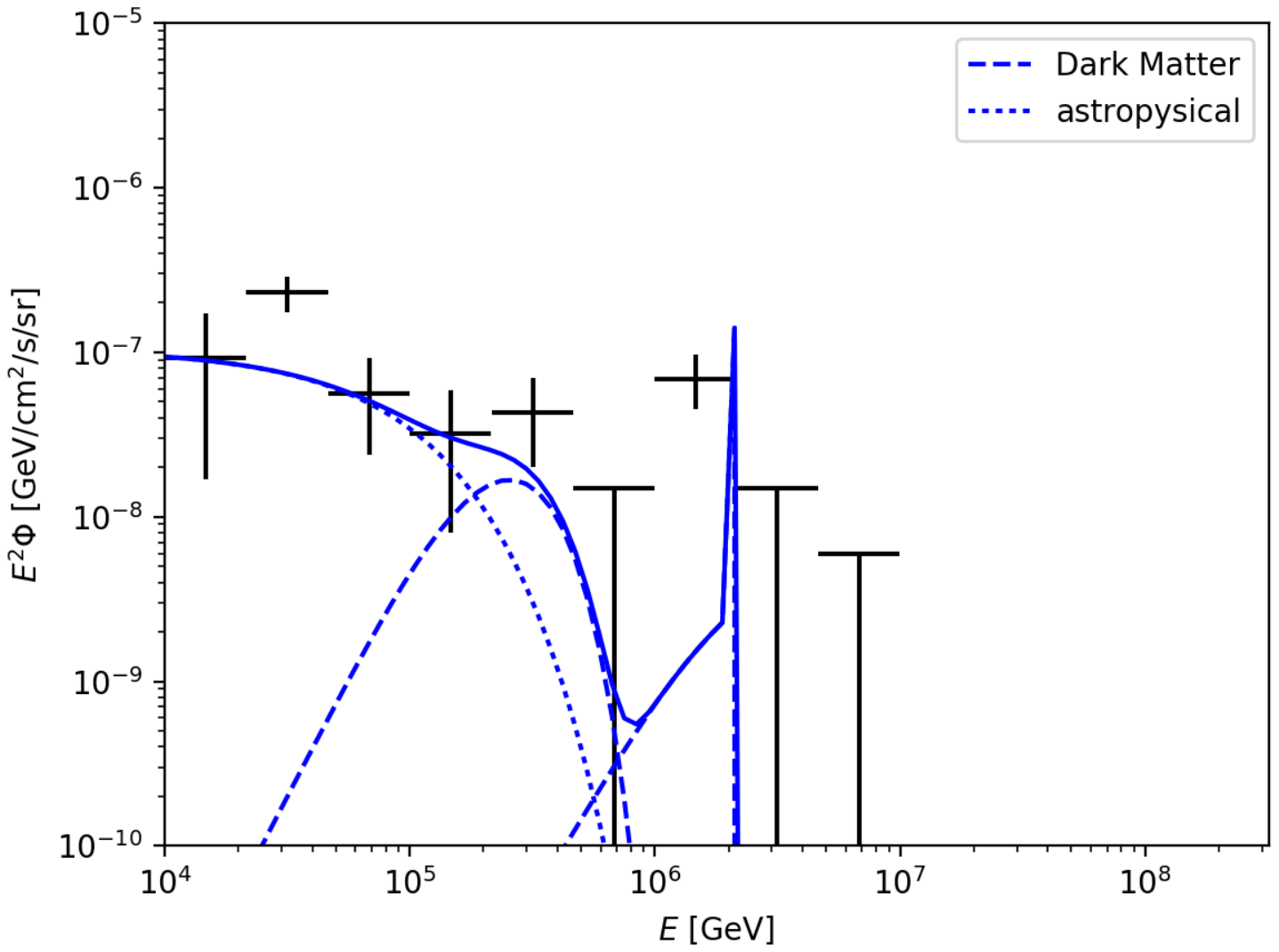}
\caption{\label{astroDM2sou} Neutrino source spectrum in Model 2a corresponds to the Fig.~\ref{astroDM2depo}.  The dashed (dotted) line corresponds to the DM (astrophysical) contributions. The solid line represents the sum of those components.}
\end{figure*}

In this model, we consider an astrophysical component for medium-energy neutrinos, which is produced by some hidden accelerators such as choked $\gamma$-ray burst jets and/or cores of active galactic nuclei~\cite{Murase:2016gly}. 
We assume that the astrophysical component has a flat energy spectrum of $E_\nu^2\Phi_\nu$ with a cutoff energy at ${\cal O}$(10) TeV. Then, we calculate a total neutrino spectrum by combining the astrophysical component with the broad and line spectra produced by the decaying DM. In Fig.~\ref{astroDM2depo} and Fig.~\ref{astroDM2sou}, we show a case that the mass of the DM is $m_\mathrm{DM}=4$~PeV, and the branching ratio into the line is BR$_\mathrm{line}=0.080$. Here we assume that the DM decays into $N\sim30$~particles with its lifetime $\tau=3.41\times10^{27}$~sec.

%

\subsubsection{Model 2b: Cutoff of the astrophysical components at  ${\cal O}$(1) PeV}

\begin{figure*}[htbp]
\centering 
\includegraphics[width=0.65\textwidth,trim=20 200 20 150, clip]{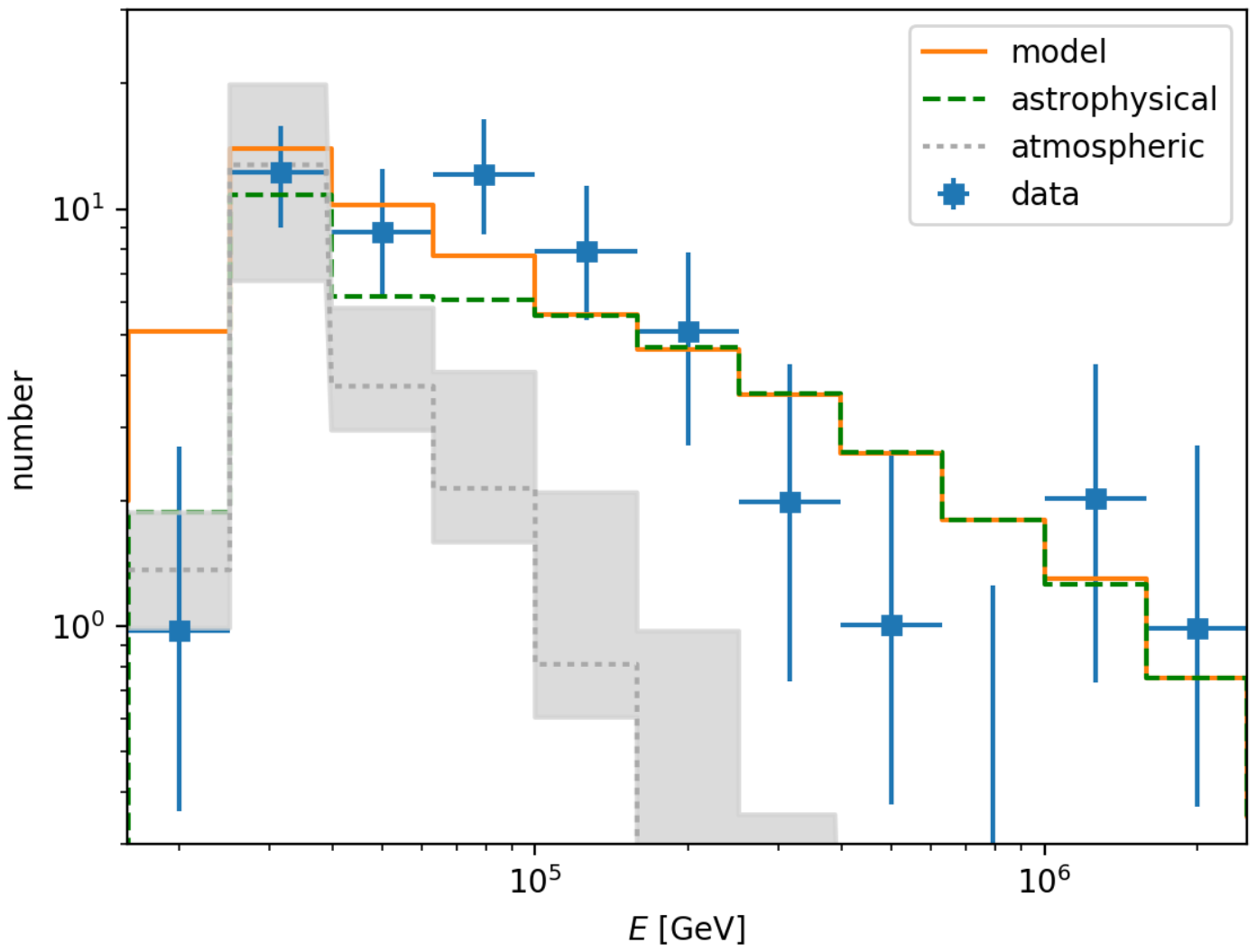}
\caption{\label{astroDMdepo1} Deposited energy histogram of the neutrino by both the decaying DM
  and the astrophysical components (Model 2b) .  Here we assume that
  the DM with its mass of $m_\mathrm{DM}=600$TeV decays into the
  $N\sim30$ particles. Contribution from the two-body decay mode is
  negligible.  In this case, the lifetime of the
  DM particle is fitted to be $\tau=5.56\times10^{26}$~s.
The normalization of the astrophysical component is same as those of \cite{Murase:2013rfa} in order.
  }
\includegraphics[width=0.65\textwidth,trim=20 200 20 150, clip]{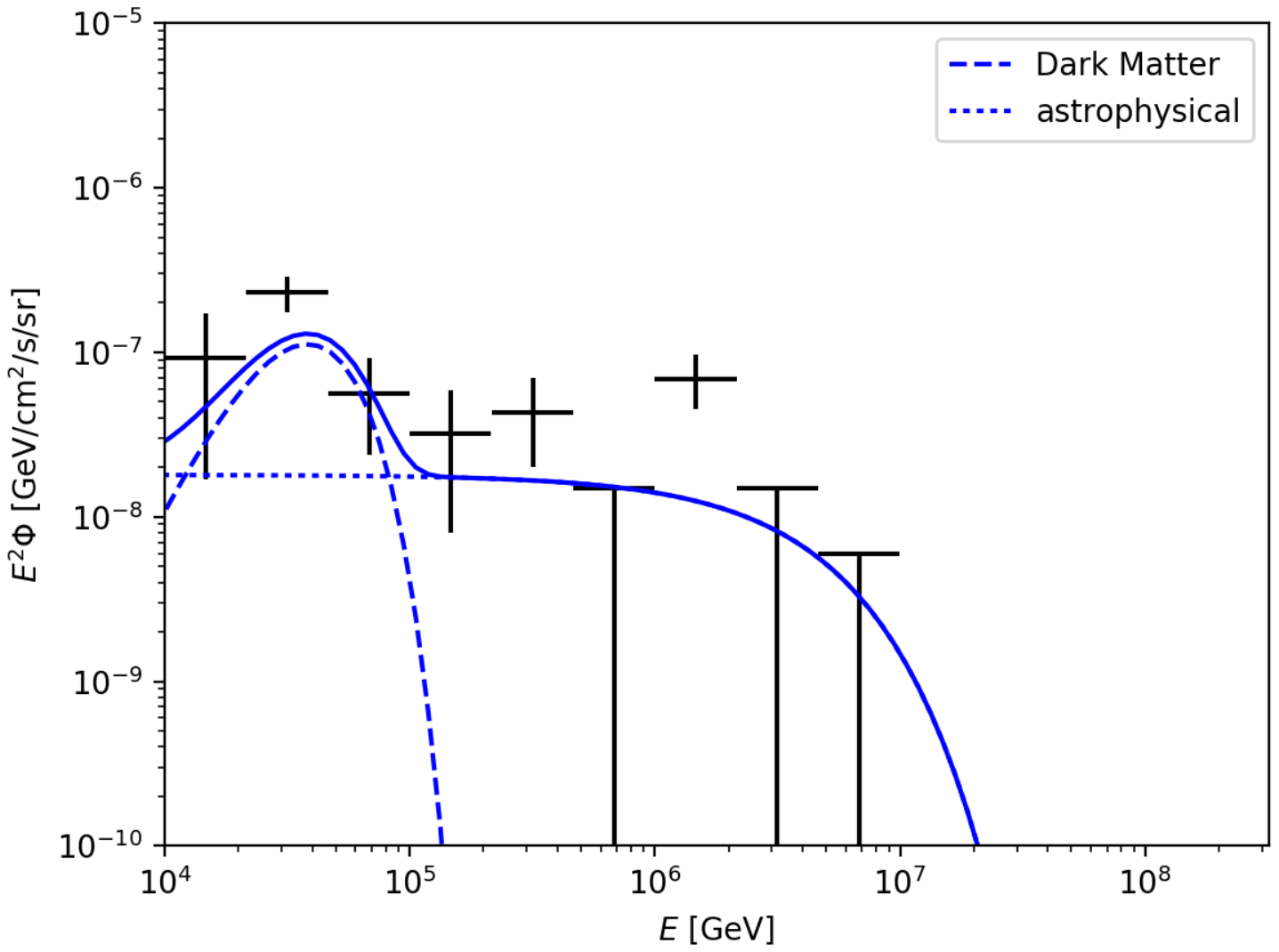}
\caption{\label{astroDMsou1} Source spectrum of the neutrino in Model 2b derived with the same parameters assumed in those of Fig.~\ref{astroDMdepo1}. Lines are the same as those in Fig.~\ref{astroDM2sou}}
\end{figure*}

The cutoff scale can be taken to be $\sim2$~PeV that is consistent with the non-observation of the Glashow resonance, and 
such a spectral suppression was predicted by plausible astrophysical models in which neutrinos are produced in galaxy clusters or starburst galaxies\cite{Murase:2016gly}.  
We show an example of the neutrino spectrum by the combination of such an astrophysical component and the decaying DM
contribution. 
The former (latter) component fits the higher (lower) energy part of the spectrum. 
In Fig.~\ref{astroDMdepo1}, and Fig.~\ref{astroDMsou1}, we plot the case that the DM particle has its
mass $m_\mathrm{DM}=600$~TeV and decays into $N\sim$30 particles.
Then the lifetime is fitted to be $\tau\sim5.56=10^{26}$~sec. Here we assume that a line spectrum
originated from the two-body decay is negligible.

\section{Note on IGRB}
\label{sec:IGRBconstraint}
The multi-body decaying DM model considered in this work is not constrained by the present data of the IGRB 
reported by the {\it Fermi} satellite, because little $\gamma$-rays are emitted. Electromagnetic emission is expected 
due to the electroweak bremsstrahlung emission accompanied by the neutrino emissions.
We have estimated the contributions to the IGRB in an analytical way. The details of our estimates are as follows,and we show that the contributions to the IGRB is about 1-10\%. 

We have proposed a DM model which decays to produce neutrinos in two- and multi- particle final states. In each branch of the decay mode, the first next order diagrams, which correspond to the electroweak bremsstrahlung, of the electroweak corrections are 
\begin{eqnarray}
\nu &\to& \nu + Z\\
&\to& \nu + q \bar{q}
\end{eqnarray}
and
\begin{eqnarray}
\nu &\to& W + \mathrm{charged}\ \ \mathrm{ lepton} (e, \mu, \tau)\\
&\to&q \bar{q} + \mathrm{charged}\ \ \mathrm{ lepton} (e, \mu, \tau).
\end{eqnarray}
The amplitudes of these corrections are almost same. We take a conservative value as the weak coupling $g\sim 0.65$, then the amplitudes are apploximately
\begin{equation}
\frac{\alpha_W}{\pi}\times\log\left(\frac{1\mathrm{PeV}}{M_W}\right)\sim{\cal O}(0.1)
\end{equation}
with $\alpha_W=g^2/(4\pi)$.

In terms of the flux contributions to the IGRB, inverse Compton emissions from those charged leptons flipped from active neutrinos produced in the electroweak corrections are the most relevant ones. 
 Since the photon spectrum have broad distributions corresponding to the energy distributions of the lepton spectra, the photon flux reduces to a few tens of percent if we look it in each energy bin. Then, we estimate that the contributions to the IGRB flux is at most 10\%. Note that this value is derived in an analytical way. 
Charged leptons are also produced in the decay of the W/Z bosons in the electroweak corrections. In this case, 
lepton spectra also have broad distributions because they are produced by the decaying mesons which have broad specturm. The reason that those mesons already have broad distributions is they are produced after the hadronizations of $q\bar{q}$ emissions due to the decays of W and/or Z. Then, the contributions of the inverse Compton emissions from those leptons to the IGRB is about a few \%.

By exactly setting up branching ratios to each modes we can quantitatively estimate the generated $\gamma$-ray flux. In this case, however,  the total amount of the $\gamma$-rays are already regulated by the branching ratios in those of analytical estimates we have shown above. For a simplified case of $\mathrm{DM}\rightarrow \nu \nu$, which gives us the most conservative estimate maximising the photon flux at a certain energy bin,  Ref.\cite{Cohen:2016uyg} numerically calculated the $\gamma$-ray emissions and showed that the contributions to the IGRB is negligible.

For contributions of the electroweak corrections to the IGRB, those $\gamma$-rays from galactic interactions are dominant. We can neglect the extragalactic contributions at TeV-PeV range since the infrared background emissions absorb the $\gamma$-rays higher than ${\cal O}$ (1)TeV.
Therefore our scenario has not been excluded by the current $\gamma$-ray observations. On the other hand, 
we may be able to detect these contributions to the IGRB by LHAASO and other future missions.
This is beyond the scope of this paper and should be a future work. 

 \section{Discussion and Conclusion}
\label{sec:conclusion}
We have built models of the multi-body decaying DM, in which decay products consist of a neutrino and ${\cal O}$(30) dark fermions. Those neutrinos produced in the decay of the DM of $m_\mathrm{DM}\sim{\cal O}$(1)PeV can contribute to the TeV-PeV neutrino spectrum. Such a model provides an example of a DM interpretation about high-energy neutrino observations without suffering from the isotropic $\gamma$-ray constraint.

In our setup, the $N^0$, which can be a DM candidate, couples to the neutral component of the lepton doublet $l$ in the standard
model. Then, $N^0$ decays in two modes: 1) one neutrino and $2n$
fermions ($S$), and 2) one line neutrino and a scalar $X$
where $X$ decays into $2n S$. In the first (second) mode, a broad
(line) spectrum of neutrino is produced. The observed spectrum should
be the sum of the neutrinos produced in these two decay modes.

We can fit the deposited energy histogram of the TeV-PeV neutrinos if
the mass of $N^0$ is $m_{\rm DM}$=4~PeV, and $N^0$ decays into $N=3n+1=31$
particles (denoted as Model 1). Then, the lifetime of $N^0$ is fitted
to be $\tau=1.46\times10^{27}$s with its branching ratio into the
line spectrum, BR$_\mathrm{line}$=0.034.  However, the spectrum
produced only by the decaying DM into ${\cal O}$(30) particles may not
completely fit the the medium-energy feature around ${\cal O}$ (10)TeV in
the plot of the neutrino energy spectrum. This indicates that we may need
an additional component, such as astrophysical contributions, to explain the whole neutrino spectrum.

Astrophysical neutrinos are assumed to conventionally have a single power-law
distribution. Although this is not generally true, the recent IceCube data indicate
that the 10-100~TeV flux may not be explained by a single astrophysical component. 
While various possibilities have been suggested, we have discussed a hybrid model 
of the astrophysical and the decaying DM contributions. 
We have studied two cases: the astrophysical components are responsible for i) the
$\sim$30~TeV excess (Model 2a), and ii) the PeV excess (Model 2b),
respectively. In Model 2a, we have assumed neutrinos from astrophysical
``hidden accelerators'' to fit the 30~TeV excess. On the other hand, the
neutrinos produced by the decaying DM contribute only to the PeV
neutrinos. On the other hand, in Model 2b, we have considered the
astrophysical neutrinos which are produced in the inelastic $pp$ scattering by
cosmic-ray protons with a flat energy spectrum. 
The decaying DM with $m_\mathrm{DM}$=600~TeV 
produces the excess feature of neutrinos at around $E_\nu$=30~TeV. 
The lifetime of the DM is fitted to be $\tau\sim10^{27}$~s in both
cases. Note that the astrophysical $\gamma$-rays produced in those models are
consistent with the IGRB observations.
Our results may imply that we need even such complex scenarios in order to explain the observation of the TeV-PeV neutrino, which shows some tension between the low-energy and high-energy data. 

Testing neutrinophillic DM models is challenging in general. One of the promising ways 
is to search for high-energy neutrino emission from nearby DM halos~\cite{Murase:2015gea}. 
If decaying DM scenarios are correct, we should detect neutrino signals from nearby galaxy clusters and galaxies with future neutrino 
telescopes such as IceCube-Gen2~\cite{Aartsen:2014njl}. Another test is to look for the spatial distribution. 
Thanks to the Galactic DM component, a slight excess around the Galactic Center is also expected~\cite{Bai:2014kba}.
Discrimination among various DM models is more difficult, but searching for inverse-Compton $\gamma$-rays by leptons from electroweak bremsstrahlung of the Galactic DM is one of the possibilities. We would need a sensitivity of $\sim 10^{-9}~{\rm GeV}~{\rm cm}^{-2}~{\rm s}^{-1}~{\rm sr}^{-1}$, which could be reached by LHASSO. 
Note that dark fermions in our model can be regarded as a boosted DM~\cite{Agashe:2014yua,Bhattacharya:2014yha,Kopp:2015bfa,Necib:2016aez,Bhattacharya:2016tma,Kong:2014mia}. However, their coupling to Standard Model particles 
are so weak that it is difficult to detect them with direct detection experiments unless additional assumptions are made. 

The model also has some cosmological implications.
In the early Universe, $S$ can be thermalized. Then, we predict a dark radiation component of the relic abundance of $S$ to be an effective number of neutrino species $N_{\rm eff} \sim 0.1$, which  could be measured by the future CMB and 21cm line
observations~\cite{Kohri:2013mxa}. The DM interpretations of the high-energy neutrinos discussed in this paper could 
be tested in near future.

\appendix
\section{A possible scenario for the particle physics model}
\label{sec:setup}

In this section, we describe a setup of particle physics models of a
strongly-interacting massive sector which includes the heavy lepton
$SU(2)$ doublet $L$, the massive scalar $X$, and the massless fermion
$S$. We assume a presence of a $Z_{2n}$ symmetry under which $S$ has a
unit charge as in Table~\ref{tab:charges}. We set the dynamical scale to
be $\Lambda \sim $~PeV.

According to the naive dimensional analysis which makes the classical
estimates and the quantum corrections the same order, any mass scales
commonly become the same order of magnitude ($\sim$ ${\cal
O}(\Lambda$)). Such a system can be described as
\begin{eqnarray}
  \label{eq:LagrangeMassive}
  {\cal L} = \frac{\Lambda^4}{(4 \pi)^2}
  f (
  L / \Lambda^{3/2},~X / \Lambda,~S/ \Lambda^{3/2},~\partial / \Lambda
)
  +{\cal L_{\rm SM}},
\end{eqnarray}
where the symbol ``$\partial$'' means the operator of derivative, and
${\cal L_{\rm SM}}$ is the Lagrangian only for the particles in the
Standard Model. Here $f$ is a generic function with taking dimensionless
arguments.

\begin{table}[t]
  \centering
  \begin{tabular}{c|ccc}
         & $S$ & $X$ & $L$ \\
    \hline
   $Z_{2n}$  &1  &$0$    &  $0$
  \end{tabular}
  \caption{Charge assignments under the $Z_{2n}$ symmetry.}
  \label{tab:charges}
\end{table}

From the scaling rule in the first term of (\ref{eq:LagrangeMassive}),
we expect the kinetic term of $X$  can be expressed by
\begin{eqnarray}
  \label{eq:kinX}
  {\cal L_{\rm kin}} \sim  \frac{\Lambda^4}{(4 \pi)^2} 
  \left(
  \frac{\partial X}{\Lambda^{2}}
  \right)^2.
\end{eqnarray}
Thus, in order to obtain the canonically-normalized kinetic term, we
see that $X$ should be multiplied by a factor of $4 \pi$,
\begin{eqnarray}
  \label{eq:scaling}
  X \to 4 \pi X
\end{eqnarray}
This scaling rule must be also applied to any fields in this sector
such as $L$ or $S$.

When we discuss the interaction term between $L$ and $\ell$, we
introduce a dimensionless parameter $\epsilon$ such as
\begin{eqnarray}
  \label{eq:L1v1}
  {\cal L}_1 \sim \epsilon 
    \frac{\Lambda^4}{(4 \pi)^2 }
  \frac{\overline{L}}{\Lambda^{3/2}}
  \frac{\ell}{\Lambda^{3/2}}
  h (
  X / \Lambda,~S/ \Lambda^{3/2}
   ),
\end{eqnarray}
with a dimensionless function $h$. Here $\epsilon$ represents the
coupling between the DM sector and $\ell$. 
The interaction term among $L$, $\ell$ and $X$, are given by
\begin{eqnarray}
  \label{eq:Lline}
  {\cal L_{\rm line}} = \epsilon 
    \frac{\Lambda^4}{(4 \pi)^2 }
  \frac{\overline{L}}{\Lambda^{3/2}}
  \frac{\ell}{\Lambda^{3/2}}
  \frac{X}{\Lambda^{}}.
\end{eqnarray}
By using the scaling rule (\ref{eq:scaling}) for $L$ and $X$, we see
that the canonically-normalized interaction term is represented by
\begin{eqnarray}
  \label{eq:LlineNorm}
  {\cal L_{\rm line}} = \epsilon 
  {\overline{L}}
  {\ell}
  {X}.
\end{eqnarray}
In the same manner, the interaction Lagrangian for the mode into a
broad spectrum is expressed by
\begin{eqnarray}
  \label{eq:Lborad}
  {\cal L_{\rm broad}} = \epsilon 
    \frac{\Lambda^4}{(4 \pi)^2 }
  \frac{\overline{L}}{\Lambda^{3/2}}
  \frac{\ell}{\Lambda^{3/2}}
  \left(
  \frac{S}{\Lambda^{3/2}}
  \right)^{2 n}.
\end{eqnarray}
Due to the scaling rule (\ref{eq:scaling}), the rescaled one is
written as
\begin{eqnarray}
  \label{eq:LbroadNorm}
  {\cal L_{\rm broad}} = \epsilon 
{
(4 \pi)^{2 n - 1} \over \Lambda^{3n - 1}
}
  {\overline{L}}
  {\ell}
  {S}^{2 n}. 
\end{eqnarray}

The $X$ and $S$ fields are also interacting each other. In total, the
Lagrangian with the canonically normalized fields is given by
\begin{eqnarray}
  \label{eq:lagrangeTot}
  {\cal L} &\sim& 
\Lambda \bar L L + 
\epsilon
{
(4 \pi)^{2 n - 1} \over \Lambda^{3n - 1}
}
 \overline{L} \ell S^{2 n}
   + \epsilon  \overline{L} \ell X 
   + \frac{(4 \pi)^{2n-1}}{\Lambda^{3n-3}} X S^{2 n}  + {\rm h.c.}
+ ... 
\end{eqnarray}
Compared with eqs.~\eqref{eq:lagrange1} and \eqref{eq:lagrange1-2}, we obtain
\begin{eqnarray}
  \label{eq:MstarEpsilon}
m_{\rm DM} \sim \Lambda,
\quad
  \epsilon \sim 
\left( 
{1 \over 4 \pi}
\right)^{2n-1}
\left(
{ \Lambda \over M_* }
\right)^{3n-1}, 
\quad
M^{3n-3} \sim 
{\Lambda^{3n-3} \over (4 \pi)^{2n-1}}.
\end{eqnarray}


\begin{acknowledgments}

This work is partially supported by JSPS KAKENHI Grant Nos.~15H03669 and 15KK0176
(RK), 26247042 and JP1701131 (KK), and MEXT KAKENHI Grant
Nos. 25105011 (RK), JP15H05889 and JP16H0877 (KK).  The work of
K.M. is supported by NSF Grant No. PHY-1620777.
\end{acknowledgments}
\bibliography{kmurase.bib}

\end{document}